# Policy based intrusion detection and response system in hierarchical WSN architecture.

Mohammad Saiful Islam Mamun[1]  A.F.M Sultanul Kabir[2], Md. Sakhawat Hossen.[2]  Md.Razib Hayat Khan[3]

[1] School of Information and System Sciences

[2] School of Information and Communication Technology

Royal Institute of Technology (KTH) Stockholm Sweden

[3] Department of Telematics

Norwegian University Of Science and Technology (NTNU), Norway

msimamun@kth.se, afmk@kth.se, hossen@kth.se, rkhan@item.ntnu.no

*Abstract*— **In recent years, wireless sensor network becomes popular both in civil and military jobs. However, security is one of the significant challenges for sensor network because of their deployment in open and unprotected environment. As cryptographic mechanism is not enough to protect sensor network from external attacks, intrusion detection system (IDS) needs to be introduced. In this paper we propose a policy based IDS for hierarchical architecture that fits the current demands and restrictions of wireless ad hoc sensor network. In this proposed IDS architecture we followed clustering mechanism to build four level hierarchical network which enhance network scalability to large geographical area and use both anomaly and misuse detection techniques for intrusion detection that concentrates on power saving of sensor nodes by distributing the responsibility of intrusion detection among different layers. We also introduce a policy based intrusion response system for hierarchical architecture.**

## I. INTRODUCTION

Wireless Sensor networks (WSN) are becoming the target point of many scientific researches. Among the technical constraints small size, limited mobility and lifetime, low battery power, storage, memory, computational power are distinguished. Such networks need to be more secured than others, as sensor nodes are more vulnerable due to operation in hostile environment. So, a number of cryptographic, communication and physical security mechanisms are proposed to fight against the vulnerabilities. Though some of the IDSs are proposed in the field of sensor networks but they are not complete in a way to build IDS and are applicable only in small geographical area. In this paper we address this problem and contribute by proposing a novel architecture for intrusion detection scheme of WSN.

## II. EXISTING CHALLENGES IN IDS ARCHITECTURE

Existing IDS architectures are not adequate to protect WSN from all kinds of inside and outside attackers. None of them are complete. Most of the existing IDSs deal with wired architecture except their wireless counterpart. Though a number of IDS architectures are proposed for ad hoc network, the architecture of WSN is even more sophisticated than wireless ad hoc network. An ideal IDS has the capability of detecting inside and outside attacks, known and unknown attacks with low false alarm rate. Existing IDS architectures that are specifically designed for sensor networks are suffering from lack of scalability and resources e.g. processing power, storage capabilities, unlimited battery backup etc.

## III. OVERVIEW OF EXISTING RESEARCH

There are few IDSs that are proposed for Wireless Ad hoc network to protect these threats. Most of them work on distributed environment which means they work on individual nodes independently and try to detect intrusion by studying abnormalities in their neighbours' behaviour. Thus they require the nodes to consume more of their processing power, battery backup, and storage space which turn IDS to be more expensive, or become infeasible for most of the applications. Some of the IDS use mobile agents in distributed environment. Mobile Agent supports sensor mobility, intelligent routing of intrusion data throughout the network, eliminating network dependency of specific nodes. But this mechanism still is not popular for IDS due to mobile agents' architectural inherited security vulnerability and heavy weight. Almost all the IDS are attack-specific which make them concentrated to some specific type of attacks. Some of them use centralized framework which make IDS capable exploiting a personal computers high processing power, huge storage capabilities and unlimited battery. On the other hand most of them work in decentralized fashion. Most of the IDS are targeted to routing layer only, but it can be enhanced to detect different types of attacks at other networking layers as well. Most of the architectures are based on anomaly detection which examines the statistical analysis of activities of nodes for detection. Some detects only intrusion while some do more like acquiring more information e.g. type of attacks, locations of the intruder etc. Though a number of IDS mechanisms are proposed Wireless ad hoc network but very few of them can be applicable for WSN because of their resource constrains. Self-Organized Criticality & Stochastic Learning based IDS [1], IDS for clustering based sensor Networks [2], A non-cooperative game approach [3], Decentralized IDS [4] are distinguished among them. Existing IDSs designed for WSN have lack of resources e.g. high processing power, high storage and battery backup etc. IDS scalability is another limitation for WSN which highly correlates with application environment, network management tool, available resource etc. We studied all the weaknesses and strong part identified of intrusion detection systems architecture to evaluate a new system. Finally, based on all the attacks and their detection system a new detection and response system is proposed which use policy based architecture that might overcome the weaknesses of the current IDS.

## IV. POLICY BASED IDS ARCHITECTURE

A policy implies predefined action pattern that is repeated by an entity whenever certain conditions occur. The architectural components of policy framework include: a Policy Enforcement Point (PEP), Policy Decision Point (PDP), and a Policy repository. The policy rules stored in Policy repository are used by PEP to implement rules or to show results. PDP translates the available data to a device-dependent format and configures the relevant PEPs. The PEP executes the logical entities that are decided by PDP [6]. Hierarchical WSN management can be realized by policy mechanism to achieve both scalability and autonomy in Intrusion detection in large geographical area. Survivability is one of the major achievements, intended from the proposed system, can be ensured by policy mechanism. So, in case of failure the system enables one component to take over the management role of another component. To achieve a policy-based management for IDS the proposed architecture has several components that evaluate policies: a Base Policy Decision Point (BPDP), a number of Policy Decision Modules (PDM).

*Base Policy Decision Point (BPDP)*

The BPDP is the controlling component of the architecture. It implements policies or intrusion rules created by the Intrusion Detection Tool (IDT) by receiving events, evaluating anomaly conditions and applying new rules, algorithms, threshold values etc.

*Policy Decision Modules (PDM)*

Policy Decision Modules are components that perform implementing sophisticated algorithms in relevant domains. Hierarchical network management that is proposed in this paper uses intermediate nodes (e.g. Regional nodes, cluster nodes) to distribute the detection tasks. Each intermediate node has its own agent called Regional or Cluster agent which collects and processes information from its domain and passes the required information to the upper layer nodes for further steps. All the intermediate nodes are also responsible to distribute command/data/message from the BPDP to lower layer agents. There is no direct communication between the member nodes.

The hierarchical architecture of policy management for WSN is shown in figure 1. It comprises of several hierarchical layers containing Intrusion Detection Agent (IDA) at each layer. Each layer represents to particular network scopes and is protected by agents affiliated to that layer. As shown in figure 1, four levels of hierarchical agents from base station to sensor nodes. They are Base Policy Decision Point (BPDP), Regional Policy Agent (RPA), Local Policy Agent (LPA), Sensor Node(S).

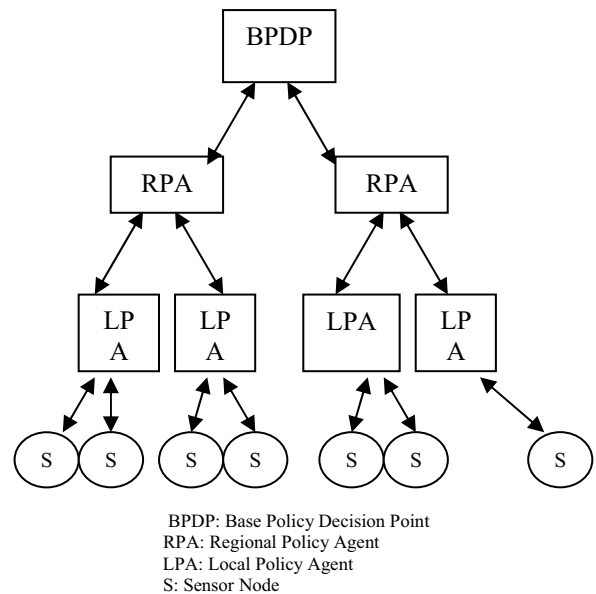

BPDP: Base Policy Decision Point
RPA: Regional Policy Agent
LPA: Local Policy Agent
S: Sensor Node

Figure 1: Hierarchical Architecture of IDS Policy Management

### A. Structure of Intrusion Detection Agent (IDA)

An IDA as shown in figure 2 consists of the following components: Pre-processor, Signature Processor, Anomaly processor and Post processor. The functionalities are described as follows:

*1) Pre-Processor*

It either collects the network traffic of the leaf level sensor when it acts as an LPA or it receives reports from lower layer IDA. Collected sensor traffic data is then abstracted to a set of variables called stimulus vector to make the network status understandable to the higher layer processor of the agent.

*2) Signature Processor*

It maintains a reference model or database called *Signature Record* of the typical known unauthorized malicious threats and high risk activities and compares the reports from the *pre-processor* against the known attack signatures. If match is not found then misuse intrusion is supposed to be detected and signature processor passes the relevant data to the next higher layer for further processing.

*3) Anomaly Processor*

It analyses the vector from the *pre-processor* to detect anomaly in network traffic. Usually statistical method or artificial intelligence is used in order to detect this kind of attack. Profile of normal activity which is propagated from Base station is stored in the database. If the activities arrived from pre-processor deviates from the normal profile in a statistically significant way, or exceeds some particular threshold value, attacks are noticed. Intrusion detection rules are basically policies which define the standard of access mechanism and uses of sensor nodes. Here database acts as a Policy Information Base (PIB) or policy repository.

*4) Post Processor*

It prepares and sends reports for the higher layer agent or base station. It can be used to display the agent status through a user interface.

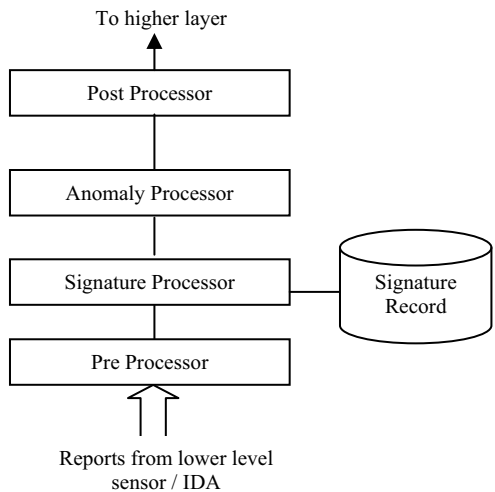

Figure 2: Intrusion Detection Agent Structure

### B. Hierarchical policy management

Except the leaf level sensor nodes all the nodes in the higher level are configured with higher energy and storage. LPA manages the sensor nodes which is more powerful than traditional sensor nodes. LPAs perform local policy-controlled configuration, filtering, monitoring, and reporting which reduces management bandwidth and computational overhead from leaf level sensor nodes to improve network performance and intrusion detection efficiency. An RPA can manage multiple LPAs. In the same way BPDP manages and controls all the RPAs. If we compare it to architectural components of policy framework then BPDP is equivalent to global PDP, RPA stands for intermediate PDP, and an LPA resembles PEP. These three level Policy management agents manage the low level sensor nodes in WSN. Policies are disseminated from the BPDP to Sensor Nodes through RPA and LPA as they are propagated from PDP to PEP in policy framework.

Policy agents described above helps the network to be reconfigured automatically to deal with fault and performance degradation according to intrusion response. One of the major architectural advantages of hierarchical structure is any node can take over the functionality of another node dynamically to ensure survivability. A flexible agent structure ensures dynamic insertion of new management functionality [6]. Proposed IDA structure ensures dynamic insertion of new rules for IDS. Dynamic update of different hierarchical agents help signature based anomaly detection effectively. So if Base station can detect any anomaly as attack it will disseminate the signature of attack to the LPA through RPA. Signature of attacks actually resembles Policy Repository of policy based architecture.

When Network administrator wants to execute some Intrusion Detection Algorithm, BPDP sends the relevant policy to all the intermediate agents. In this design no intrusion detection modules are stored in leaf level Sensor nodes in order to preserve their valuable resources. On the other hand each intermediate agent is kept on standby mode which improves the real-time performance effectively.

## V. INTRUSION DETECTION ENTITIES

In this architecture, total area of sensor network will be divided into several regions (e.g. GSM cells) [11]. Sensor nodes in each region are monitored by a cluster node. Two or more cluster nodes will be monitored by a regional node. In turn, Regional nodes will be controlled and monitored by the Base station. Figure 3 shows an example of the model. It is noted that each region is not necessarily bound to be strictly hexagonal.

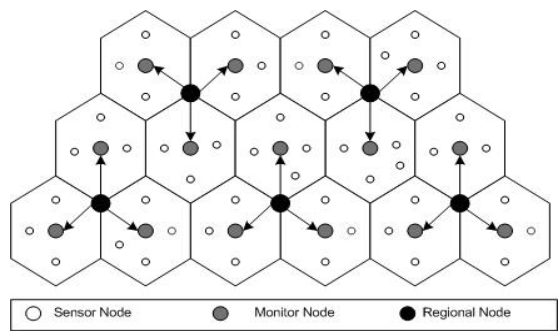

Figure 3: Intrusion detection entities

*Sensor Nodes* have two functionality: Sensing and Routing. Each of the sensor nodes will sense the environment and exchange data in between sensor nodes and cluster node. As sensor nodes have much resource constraints, in this model, there is no IDS module installed in the leaf level sensor nodes.

*Cluster Node* plays as a monitor node for the sensor nodes where LPA works. One cluster node is assigned for each of the small region e.g. GSM cell. It will receive data from sensor nodes, analyses and aggregates information to send to the regional node. It is more powerful than sensor nodes and has intrusion detection capability built into it.

*Regional Node* will monitor and receive the data from neighbouring cluster heads where RPA works. It sends the combined alarm to the upper layer base station. It is also a monitor node with all IDS functionalities. If thousands of sensor nodes are placed at the leaf level then whole area will be split into several regions to make the sensor network more scalable and more manageable.

*Base Station* is the topmost part of architecture empowered with human support. BPDP works here. It will receive information from Regional nodes and distribute the information to the users on demand.

## VI. INTRUSION RESPONSE

There are differences between intrusion detection and intrusion prevention. If a system has intrusion prevention, it is assumed that intrusion detection is built in. IDSs are designed to welcome intrusion to get into system; where as Intrusion Prevention System (IPS) actually attempts to prevent access to the system from the very beginning. IPS operates similar to IDS with one critical difference: "IPS can block the attack itself; while an IDS sits outside the line of traffic and observes, an IPS sits directly in line of network traffic. Any traffic the IPS identifies as malicious is prevented from entering the network [7]." So in case of IDS "Intrusion Response" should be the right title for recovery.

There are two different approaches for intrusion response: Hot response or Policy based response [8]. *Hot response* reacts by launching local action on the target machine to end process, or on the target network component to block traffic. E.g. kill any process, Reset connection etc. It does not prevent the occurrence of the attack in future. On the other hand, *Policy based response* works on more general scope. It considers the threats reported in the alert, constraints and objectives of the network. It modifies or creates new rules in the policy repository to prevent an attack in the future. In case of reaction approaches countermeasure could have negative impact on the network. For example, an administrator might not prefer to react, as the risk of the detected attack is less than the risk resulting from triggering off a candidate countermeasure [10]. So a risk assessment method is needed to evaluate and justify the risk of an attack and its countermeasure. In our proposed IDS, we allow policy based response system. Base station's Policy decision point and other policy decision modules take part in the response mechanism together. Intrusion can be detected either in Cluster node or Regional node. Finally, base station can be involved anytime if network administrator wants to upgrade signature database or policy stored in intermediate agent. Intrusions are detected automatically according to the policy implemented by BPDP. Re-action is also automatic but administrator may re-design the architecture according to special requirement.

In [9] a novel intrusion detection and response system is implemented. We have applied their idea in our response mechanism with some modification. Our IDS system considers each sensor nodes into one of five classes: *Fresh*, *Member*, *Unstable*, *Suspect* or *Malicious*. We have Local Policy Agent, Regional Policy Agent and finally Base Policy Decision Point to take decision about the sensor node's class placement. Routeguard mechanism use *Pathrating* algorithm to keep any node within these five classes [9]. In our model, we have policy or rules defined in Base station's BPDP to select any node to be within these five classes as shown in figure 4. When a new node is arrived, it will be classified as *Fresh*. For a pre-selected period of time this new node will be in *Fresh* state. By this time LPA will check whether this node is misbehaving or not. After pre-defined time it will enter into Member state automatically if no misbehave is detected. Otherwise the node's classification will be changed to *Suspect* state. In *Member* state nodes are allowed to create, send, receive or forward packets. In this time Member nodes are monitored by Watchdog at LPA in Cluster node. If the node misbehaves, its state will be changed to *Unstable* for short span of time. During *Unstable* state nodes will be kept under close observation of LPA. If it behaves well then it will be transferred to *Member* state again. A node in *Unstable* state will be converted to *Suspect* state in two cases: Either the node was in *Unstable* state and interchanged its state between *Member* and *Unstable* state for a particular times (threshold value defined in LPA) within a predefined period, or the node was misbehaving for long time (threshold value). LPA's Post processor sends "red alert" to RPA whenever *Suspect* node is encountered. Then the suspected node will be completely isolated from the network. It will not be allowed to send, receive, or forward packets and will be banned temporarily for short time. Any packets received from suspected node are simply discarded. After a certain period of time the node will be reconnected and monitored closely for extensive period of time by three layer agents. If watchdogs report well then node status will be changed to *Unstable*. However, if it continues misbehaving then it will be labelled as *Malicious* and banned permanently from the network. To ensure that this malicious node will never try to reconnect, its MAC address or any unique ID will be added to *Signature Record Database* of LPA and all other higher layer agents.

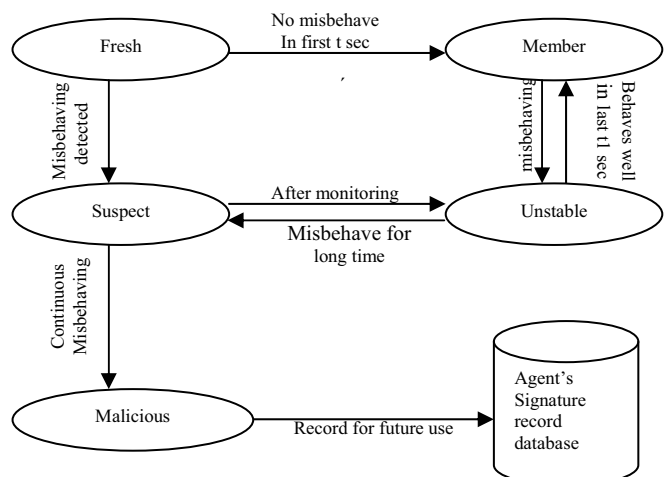

Figure 4: Operation of Intrusion Response

Survivability is one of the major factors that are predicted from every system. We consider base stations to be failure free. But the Regional nodes or cluster nodes may be unreachable due to failure or battery exhaustion. So, in case of failures or any physical damage of Regional nodes or Cluster nodes, control of that node should be taken over by another stable node. So in our proposed architecture if any Regional node fails, then its control will be shifted to the neighbour Regional node dynamically.

So, control of the Cluster nodes and sensor nodes belonging to that Regional node will be shifted automatically to the neighbour node. In the same way if any cluster node fails then control of that cluster node will be transferred to the neighbour Cluster node.

So, if any LPA is unreachable due to failure or battery exhaustion of cluster nodes, neighbour LPA will take the charges of leaf level sensor nodes which was in the area of fault cluster node. In the same way due to Regional nodes' failure neighbour Regional node's RPA will take over the functionality of all the cluster node's LPA and sensor nodes belonged to the faulty Regional node dynamically.

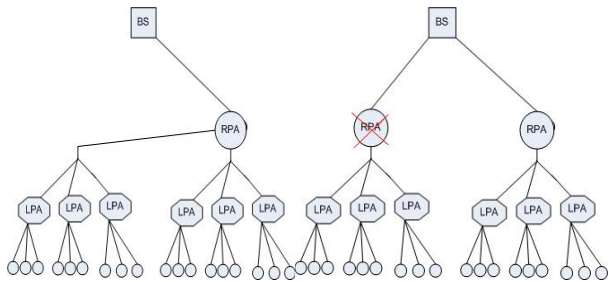

Figure 5 : Regional node's failure

As we mentioned, Cluster nodes or regional nodes have no direct communications between them. So how will Cluster node or Regional node determine about the failure of its neighbour? Actually in the proposed architecture Base station has direct or indirect connections with all its leaf nodes. Base station has direct connection with Regional node. So if any Regional node fails Base station can identify the problem and select one of its neighbour nodes dynamically according to some predefined rule in BPDP. Then BPDP needs to supply the policy, rules, or signatures database to the selected new neighbour Regional node. In the same way if any cluster node fails then policy, rules or signatures of the failed cluster node will be supplied by the BPDP through relevant RPA. As Base station is much more powerful node with large storage; all the signature databases, anomaly detection rules or policies are stored primarily as backup in Base station. This back up system increases reliability of the whole network system.

## VII. FUTURE WORK

Proposed IDS system is highly extensible, in that as new attack or attack pattern are identified, new detection algorithm can be incorporated to policy. Possible venues for future works include:

- Present model can be extended by exploring the secure communication between base station, Regional node and cluster node.
- Implementation of Risk Assessment System in the manager stations to improve the reaction capability of IDS.
- We actually focus on the general idea of architectural design for IDS and how a policy management system can be aggregated to the system. But an extensive work needs to be done to define Detection and Response policy as well.
- Overall, more comprehensive research is needed to measure the current efficiency of IDS, in terms of resources and policy, so that improvements of its future version(s) are possible.
- Building our own Simulator: all the previous research were based on three layer architecture, so we are planning to create our own simulator that will simulate our four layer policy based architecture.

## VIII. CONCLUSION

In this paper, we propose a novel architecture of IDS for ad hoc sensor network based on hierarchical overlay design. We propose a response mechanism also according to proposed architecture. Our design of IDS improves on other related designs in the way it distributes the total task of detecting intrusion. Our model decouples the total work of intrusion detection into a four level hierarchy which results in a highly energy saving structure.

Policy based mechanism is a powerful approach to automating network management. The management system for intrusion detection and response system described in this paper shows that a well structured reduction in management traffic can be achievable by policy management. This policy-based architecture upgrades adaptability and re-configurability of network management system which has a good practical research value for large geographically distributed network environment.